\documentclass[preprint2]{aastex}

\slugcomment{To appear in the Astrophysical Journal}

\shorttitle{Center-to-Limb Variation}
\shortauthors{Koesterke, Allende Prieto, Lambert}

\begin{document}


\title{Center-to-Limb Variation of Solar 3-D Hydrodynamical Simulations}


\author{L. Koesterke, C. Allende Prieto\footnote{Present address: Mullard Space Science Laboratory, University College London, Holmbury St. Mary, Surrey, RH5 6NT, UK}
and D. L. Lambert}
\affil{McDonald Observatory and Department of Astronomy, 
University of Texas, Austin, TX 78712 USA}


\begin{abstract}
We examine closely the solar Center-to-Limb variation of continua and
lines and compare observations with predictions from both a 3-D
hydrodynamic simulation of the solar surface (provided by M. Asplund and
collaborators) and 1-D model atmospheres. Intensities from the 3-D time
series are derived by means of the new synthesis code {\sc
Ass$\epsilon$t}, 
which overcomes
limitations of previously available codes by including a
consistent treatment of scattering and allowing for arbitrarily complex
line and continuum opacities. In the continuum, we find very similar
discrepancies between synthesis and observation for both types of model
atmospheres. This is in contrast to previous studies that used a
``horizontally'' and time averaged representation of the 3-D model 
and found a significantly larger disagreement with observations.
The presence of temperature and velocity fields in the 3-D simulation provides
 a significant advantage when it comes to reproduce solar spectral line shapes.
Nonetheless, a comparison of observed and synthetic equivalent widths
reveals that the 3-D model also predicts more uniform abundances as a 
function of position angle on the disk. We conclude that the 3-D simulation 
provides not only
a more realistic description of the gas dynamics, but, despite its simplified
treatment of the radiation transport, it also predicts reasonably well the
observed Center-to-Limb variation, which is indicative of a thermal structure 
 free from significant systematic errors.
\end{abstract}


\keywords{
hydrodynamics --- line: formation --- radiative transfer --- Sun:
abundances --- Sun: photosphere
}



\section{Introduction}

A few years back, it was realized that one of the most 'trusted'
absorption lines to gauge the oxygen abundance in the solar photosphere, the
forbidden [OI] line at $\lambda$6300, was blended with a Ni I transition.
 These two transitions overlap so closely that only a
minor distortion is apparent in the observed feature. Disentangling the
two contributions with the help of a 3-D hydrodynamical simulation of
surface convection led us to propose a reduction of the solar
photospheric abundance by $\sim$ 30\% \citep{allIII}. Using the same solar
model, subsequent analysis of other atomic oxygen and OH lines confirmed
the lower abundance, resulting in an average value $\log
\epsilon$\footnote{$\epsilon({\rm X})= {\rm N(X)/N(H)}\!\cdot\!10^{12}$}
(O)$= 8.66 \pm 0.05$ \citep{asp04}.

This reduction in the solar O/H ratio, together with a parallel downward
revision for carbon \citep{all02, asp05b}, ruins the nearly perfect
agreement between models of the solar
interior and seismological observations \citep{bah05, del06, lin07}. A
brief overview of the proposed solutions is given by \citet{all07}.
Interior and surface models appear to describe two different stars.

Supporters of the new hydrodynamical models, and the revised surface
abundances, focus on their strengths: they include more
realistic physics, and are able to reproduce extremely well 
detailed observations (oscillations, spectral line asymmetries and net
wavelength blueshifts, granulation contrast and topology). Detractors
emphasize the fact that the new models necessarily employ a simplified
description of the radiation field and they have not been
tested to the same extent as classical 1-D models. The
calculation of spectra for 3-D time-dependent models is a demanding
task, which is likely the main reason why some fundamental tests have not yet
been performed for the new models.

On the basis of 1-D radiative transfer calculations, \citet{ayr06}
suggest that the thermal profile of the solar surface convection
simulation of Asplund et al. (2000) may be incorrect. \citet{ayr06} make
use of a 1-D average, both 'horizontal' and over time, of the 3-D
simulation to analyze the center-to-limb variation in the continuum,
finding that the averaged model performs much more poorly than the
semi-empirical FAL C model of Fontenla et al. (1993). When the FAL C
model is adopted, an analysis of CO lines leads to a much higher oxygen
abundance, and therefore \citet{ayr06} question the downward revision
proposed earlier.

\citet{asp05b} argue that when classical 1-D model atmospheres are
employed, the inferred oxygen abundance from atomic features differs
by only 0.05 dex between an analysis in 1-D and 3-D. The difference is
even smaller for atomic carbon lines.  When the hydrodynamical model is
considered, there is good agreement between the oxygen abundance
inferred from atomic lines and from OH transitions
\citep{asp04, sco06}. 
A high value of the oxygen abundance is
derived only when considering molecular tracers in one dimensional
atmospheres, perhaps not a surprising result given the high sensitivity
to temperature 
of the molecular dissociation. A low oxygen abundance ($\log
\epsilon$(O) $= 8.63$) value is also deduced from atomic lines and
atmospheric models based on the inversion of spatially resolved
polarimetric data \citep{soc07}. 

Despite the balance seems favorable to the 3-D models and the low values 
of the oxygen and carbon abundances, a failure of the 3-D model to
match the observed limb darkening, as suggested by the experiments of
\citet{ayr06}, would be reason for serious concern. In the present
paper, we perform spectral synthesis on the solar surface convection
simulation of Asplund et al. (2000) with the goal of testing its ability
to reproduce the observed center-to-limb variations of both 
the continuum intensities, and the equivalent widths of spectral lines. 
We compare its performance with commonly-used theoretical 
1-D model atmospheres. Our calculations are rigorous:
they take into account the four-dimensionality of the hydrodynamical
simulation: its 3-D geometry and time dependency. 
After a concise description of our
calculations in Section 2, \S 3 outlines the comparison with solar
observations and \S 4 summarizes our conclusions.

\section{Models and Spectrum Synthesis}

We investigate the Center-to-Limb Variation (CLV) of the solar spectrum
for the continuum and lines. Snapshots taken from 3-D
hydrodynamical simulations of the solar surface by \citet{asp00} serve
as model atmospheres. The synthetic continuum intensities and line
profiles are calculated by means of the new spectrum synthesis code {\sc
Ass$\epsilon$t} (Advanced Spectrum Synthesis 3-D Tool), which is
designed to solve accurately the equation of radiation transfer in
3-D. The new synthesis code will be described in detail by \citet{koeI}
and only the key features are highlighted in subsequent sections.

\subsection{Hydrodynamic Models}

The simulation of solar granulation was carried out with a 3-D,
time-dependent, compressible, radiative-hydrodynamics code
\citep{nor90,ste89,asp99}. The simulation describes a volume of
6.0\,x\,6.0\,x\,3.8\,Mm (about 1\,Mm being above $\tau_{\rm cont}
\approx 1$) with 200\,x\,200\,x\,82 equidistantly spaced grid points
over two hours of solar time. About 10 granules are included in the
computed domain at any given time.

99 snapshots were taken in 30\,s intervals from a shorter sequence of
50\,min. The grid points and the physical dimensions are changed to
accommodate the spectrum synthesis: The horizontal sampling is reduced by
omitting 3 out of 4 grid points in both directions; the vertical
extension is decreased by omitting layers below $\tau_{\rm cont}^{\rm
min} \approx 300$ while keeping the number of grid points in
$z$-direction constant, i.e. by increasing the vertical sampling and
introducing a non-equidistant vertical grid. After
these changes, a single snapshot covers approximately a volume of
6.0\,x\,6.0\,x\,1.7\,Mm with 50\,x\,50\,x\,82 grid points \citep{asp00}.

\subsection{Spectrum Synthesis}

Compared to the spectrum synthesis in one dimension, the calculation of
emergent fluxes and intensities from 3-D snapshots is a tremendous task, 
even when LTE is applied. Previous investigations (e.g., \citet{asp00, lud07})
were limited to the calculation of a single line
profile or a blend of very few individual lines on top of 
constant background opacities, and without scattering. In order to overcome
these limitations, we devise a new scheme that is capable of dealing
with arbitrary line blends, frequency dependent continuum opacities, and
scattering. The spectrum synthesis is divided into five separate tasks
that are outlined below.  A more detailed description which contains all
essential numerical tests will be given by \citet{koeI}.

\subsubsection{Opacity Interpolation}
\label{sss_opai}

For the 3D calculations we face a situation in which we
have to provide detailed opacities for $\approx 2\!\cdot\!10^7$ grid
points for every single frequency under consideration. 
Under the assumption of LTE, the size of the problem can be reduced
substantially by using an interpolation scheme to derive
opacities from a dataset
that has orders-of-magnitude fewer datapoints.  We introduce an {\em
opacity} grid that covers all grid points of the snapshots in the
temperature-density plane. The grid points are regularly spaced in
log\,$T$ and log\,$\rho$ with typical intervals of 0.018\,dex and
0.25\,dex, respectively.

We use piecewise cubic Bezier
polynomials that do not introduce artificial extrema \citep{aue03}. To
enable 3$^{\rm rd}$-order interpolations close to the edges, additional
points are added to the opacity grid. The estimated interpolation error
is well below 0.1\% for the setup used throughout the present paper.

\subsubsection{Opacity Calculation}
\label{sss_opac}

We use a modified version of {\sc SYNSPEC} \citep{hub95} to prepare 
frequency-dependent opacities for the relatively small numbers of grid
points in the {\em opacity} grid. The modifications allow for the
calculation of opacities on equidistant log($\lambda$) scales, to output
the opacities to binary files, and to skip the calculation of intensities.

Two datasets are produced. Continuum opacities are calculated at
intervals of about 1\,\AA\ at 3000\,\AA. Full opacities (continuum and 
lines) are provided at a much finer spacing of $0.3\,v_{\rm min}$, with
$v_{\rm min}$ being the thermal velocity of an iron atom at the minimum
temperature of all grid points in all snapshots under consideration. A
typical step in wavelength is $2.7\!\cdot\!10^{-3}$\,\AA\ at 3000\,\AA, 
which corresponds to 0.27 km\,s$^{-1}$. 

We adopt the solar photospheric abundances recently proposed by 
\citet{asp05a}, with carbon and oxygen abundances of ${\rm
log}\,\epsilon = 8.39$ and 8.66, respectively, which are about 30\% less
than in earlier compilations \citep{gre98}.  We account for bound-bound and
bound-free opacities from hydrogen and from the first two ionization
stages of He, C, N, O, Na, Mg, Al, Si, Ca and Fe. Bound-free cross
sections for all metals but iron are taken from {\sc TOPBASE} and
smoothed as described by \citet{allI}. Iron bound-free opacities are
derived from the photoionization cross-sections computed by the 
Iron Project (see, e.g., \citet{nah95, bau97}), after smoothing.

Bound-bound ${\rm log}\,(gf)$ values are taken from Kurucz, augmented by
damping constants from \citet{bar00} where available.  We also account
for bound-free opacities from H$^{-}$, H$_2^+$, CH and OH, and for a few million
molecular lines from the nine most prominent molecules in the wavelength
range from 2200\,\AA\ to 7200\,\AA. Thomson and Rayleigh (atomic
hydrogen) scattering are considered as well, as described below in \S
\ref{sss_scat}. The equation of state is solved considering the first 99
elements and 338 molecular species. Chemical equilibrium is assumed for
the calculation of the molecular abundances, and the atomic abundances
are updated accordingly (private comm. from I. Hubeny).

\subsubsection{Scattering}
\label{sss_scat}

We employ a background approximation, calculating 
the radiation field $J_\nu$ for the sparse continuum frequency points
for which we have calculated the continuum opacity without any
contribution from spectral lines. The
calculation starts at the bluemost frequency and the velocity field is
neglected at this point: no frequency coupling is present. 
The opacities for individual grid
points are derived by interpolation from the {\em opacity} grid, and the
emissivities are calculated assuming LTE. As mentioned above, we 
include electron (Thomson) scattering and 
Rayleigh scattering by atomic hydrogen. An Accelerated Lambda
Iteration (ALI) scheme is used 
to obtain a consistent solution of the mean radiation
field $J_\nu$ and the source function $S_\nu$ at all grid points. In
turn, $J_\nu$ is calculated from $S_\nu$ and vice versa, 
accelerating the iteration by amplifying $\Delta J_\nu = J_\nu^{\rm New}
- J_\nu^{\rm Old}$ by the factor $1\,/\,(1-\Lambda^*)$ with $\Lambda^*$
being the approximate lambda operator \citep{ols87}. Generally the mean
radiation from the last frequency point, i.e. the frequency to the blue,
serves as an initial guess of $J_\nu$ at the actual frequency. At the
first frequency point, the iteration starts with $J_\nu=S_\nu$.

The formal solution, i.e. the solution of the equation
$J_\nu=\Lambda\,S_\nu$, is obtained by means of a short characteristics
scheme \citep{ols87}. For all grid points the angle-dependent
intensity $I_\nu^\mu$ is derived by integrating the source function
along the ray between the grid point itself and the closest intersection
of the ray with a horizontal or vertical plane in the mesh. The operator
$\Lambda^*$ needed for the acceleration is calculated within
the formal solution.

For the present calculations, $J_\nu$ is integrated from $I_\nu^\mu$ at
48 angles (6 in $\mu$, 8 in $\phi$). The integration in $\mu$ is
performed by a three-point Gaussian quadrature for each half-space,
i.e. for rays pointing to the outer and the inner boundary,
respectively. The integration in $\phi$ is trapezoidal. 
The opacities and source functions are
assumed to vary linearly (1$^{\rm st}$-order scheme) along the ray. 

In order to integrate the intensity between the grid point and the point
of intersection where the ray leaves the grid cell, the opacity, source
function and the specific intensity ($\kappa_\nu$, $S_\nu$, $I_\nu^\mu$)
have to be provided at both ends of the ray. Since the point where the
ray leaves the cell is generally not a grid point itself, an
interpolation scheme has to be employed to derive the required
quantities. We perform interpolations in two dimensions on the surfaces of
the cuboids applying again Bezier polynomials with control
values that avoid the introduction of artificial extrema. 
The interpolation may introduce a noticeable source of numerical
inaccuracies. Detailed tests, using an artificial 3-D structure 
constructed by horizontally replicating a 1-D model, revealed that
a 3$^{\rm rd}$-order interpolation scheme provides sufficient
accuracy where linear interpolations fail in reproducing the radiation field:
the mean relative errors are 0.5\% and 0.05\% for linear and cubic 
interpolation, respectively.

It is possible, in terms of computing time, to calculate $J_\nu$
from the full opacity dataset for all frequencies (our 'fine' sampling).
However, since the total effect of scattering for the 
solar case in the optical is quite small, the differences between the
two methods are negligible. Therefore, we apply the faster method
throughout this paper. Note that in both approximations (using
background or full opacities), the calculation of the
mean radiation field does not account for any frequency coupling.

\subsubsection{Calculation of Intensities and Fluxes}
\label{ss_cif}

The emergent flux is calculated from the opacities of the full dataset
provided at the fine frequency grid. Again, the opacities for
individual grid points are derived by interpolation from the {\em
opacity} grid and the emissivities are calculated from LTE. The mean
background radiation field $J_\nu$ is interpolated from the coarser
continuum frequency grid to the actual frequency, and it contributes to the
source function at all grid points via Thomson and Rayleigh (atomic
hydrogen) scattering opacities.

The integration along a ray is performed in the observer's frame by following
long characteristics from the top layer down to optical depths of 
$\tau_{\rm Ray}> 20$. Frequency shifts due to the velocity field are applied to the
opacities and source functions. Each ray starts at a grid point of the
top layer and is built by the points of intersection of the ray and the
mesh. At these points of intersection an interpolation in
three dimensions is generally performed, i.e. a 2-D geometric interpolation in the
X-Y, X-Z or Y-Z-plane, respectively, is enhanced by an interpolation in
frequency necessitated by the presence of the velocity field. Additional
points are inserted into the ray to ensure full frequency coverage of
the opacities. This
is done when the difference of the velocity field projected onto the ray
between the entry and exit point of a grid cell exceeds the frequency
spacing of the opacity. Without these additional points and in the
presence of large velocity gradients, line opacities could be underestimated
along the ray --a line could be shifted to one side at the
entry point and to the other side at the exit point--, leaving only
neighboring continuum opacities visible to both points while the line is
hidden within the cell.

Similar to the calculation of the mean radiation field $J$ described in
Sect.~\ref{sss_scat}, all interpolations in both space and frequency are
based on piecewise cubic Bezier polynomials. 
It is not completely trivial to mention that for the
accurate calculation of the emergent intensities, the application of a
high-order interpolation scheme is much more important than it is for
the calculation of the mean background radiation field
(Sect.~\ref{sss_scat}). Here we are calculating precisely the
quantity we are interested in, i.e. specific intensities. But, in
addition to that, we deal with interpolations in three dimensions (2-D in
space, 1-D in frequency) instead of a 2-D interpolation in space. Hence,
any quantity is derived from 21  1-D interpolations rather than
just 5.

In the standard setup of the 3-D calculations, 20 rays are used for the
integration of the flux $F_\nu$ from the intensities
$I_\nu^\mu$. Similar to the integration of $J$ described in
Sect.~\ref{sss_scat}, the integration in $\mu$ is a three-point Gaussian
quadrature, while the integration in $\phi$ is trapezoidal. Eight
angles in $\phi$ are assigned to the first two of the $\mu$ angles while
the last and most inclined angle with the by far smallest (flux)
integration weight has 4 contributing $\phi$ angles. Note that for the
investigation of the Center-to-Limb variation, the number of angles and
their distribution in $\mu$ and $\phi$ differs considerably from this
standard setup, as explained below (Sect.~\ref{ss_continuum}).

\subsection{Spectra in 1-D}
\label{ss_1d}

To facilitate consistent comparisons of spectra from 3-D {\em and} 1-D
models, the new spectrum synthesis code {\sc Ass$\epsilon$t} accepts
also 1-D structures as input. Consistency is achieved by the
use of the same opacity data (cf. Sect.~\ref{sss_opac}) and its
interpolation (if desired in 1-D, cf. Sect.~\ref{sss_opai}) and by the
application of the same radiation transfer solvers, i.e. 1$^{\rm
st}$-order short and long-characteristic schemes
(cf. Sect.~\ref{sss_scat} and \ref{ss_cif}, respectively). All angle
integrations are performed by means of a three-point Gaussian
formula. This leaves the interpolations inherent to the radiation
transfer scheme in 3-D as the only major inconsistency between the
spectra in 1-D and 3-D. Numerical tests have revealed that these 
remaining inconsistencies are quite small, as we will report in an 
upcoming paper.

\subsection{Solar Model}
\label{ss_km}

Our choice is not to use a semi-empirical model of the solar atmosphere
as a 1-D comparison with the 3-D hydrodynamical simulation, but a
theoretical model atmosphere. Semi-empirical models take
advantage of observations to constrain the atmospheric structure, a fact
that would constitute an unfair advantage over the 3-D simulation. Some
semi-empirical models, in particular, use observed limb darkening
curves, and of course it is meaningless to test their ability to
reproduce the same or different observations of the center-to-limb
variation in the continuum. Consequently we are using models from
Kurucz, the MARCS group, and a horizontal- and time-averaged
representation of the 3-D hydrodynamical simulations. 
 
We have derived a 1-D solar reference model from the Kurucz grid
 \citep{kur93}. The reference model is derived from
3$^{\rm rd}$-order interpolations in $\tau$, $T_{\rm eff}$, ${\rm
log}\,g$, and $Z$. Details of the interpolation scheme will be presented
elsewhere. We have adopted the usual values of
$T_{\rm eff}=5777\,{\rm K}$ and ${\rm log}\,g=4.437$ (cgs) but a reduced 
metallicity of ${\rm log}\,(Z/Z_\sun)=-0.2$ in an attempt to account
globally for the difference between the solar abundances (mainly iron) 
used in the calculation of the model and more recent values, 
as described by \citet{all06}. To
avoid a biased result by using a single 1-D comparison model, we have 
also experimented with a solar MARCS model kindly provided by M. Asplund, 
and a solar model interpolated from the more recent ODFNEW grid from 
Kurucz (available on his website\footnote{http://kurucz.harvard.edu/}).
No metallicity correction was applied to these newer solar models.

In earlier investigations \citep{ayr06}, a 1-D representation
\citep{asp05b} of the 3-D time series, i.e. a 
'horizontal' average\footnote{
Horizontal average, in this context, refers to the mean value
over  a  surface with constant vertical optical depth, rather than over
a constant geometrical depth.
}
over time, has been used to study the thermal profile of the 3-D model. 
While this approximation allows easy handling by means of a 1-D radiation
transfer code, the validity of this approach has never been
established. In order to investigate the limitations of this 
shortcut,  we compare its Center-to-Limb variation in the continuum  
with the exact result from the 3-D radiation transfer on the full 
series of snapshots.

\section{Center-to-Limb Variation}

\subsection{Continuum}
\label{ss_continuum}

\citet{nec94, nec03, nec05} investigated the Center-to-Limb variation of the Sun
based on observations taken at the National Solar Observatory/Kitt Peak
in 1986 and 1987. They describe the observed intensities across the
solar disk as a function of the heliocentric distance by 5$^{\rm
th}$-order polynomials for 30 frequencies between 303\,nm and
1099\,nm. Similar observations by \citet{pet84} and \citet{els07}
(with a smaller spectral coverage)
indicate that \citet{nec94} may have overcorrected for scattered light, 
but confirm a level of accuracy of $\approx$0.4\%. 
We have calculated fluxes and intensities for small spectral regions
($\pm5\,{\rm km\,s^{-1}}$) around eight frequencies
(corresponding to standard air wavelengths of 
3033.27\,\AA, 3499.47\,\AA, 4163.19\,\AA, 4774.27\,\AA, 5798.80\,\AA,
7487.10\,\AA, 8117.60\,\AA, 8696.00\,\AA) and compare monochromatic
synthetic intensities with the data from \citet{nec94}. Because the
spectral regions are essentially 
free from absorption lines, the width of the bandpass
of the observations varying between 1.5 km\,s$^{-1}$ in the blue
(3030\,\AA) and 1.9 km\,s$^{-1}$ in the red (10990\,\AA) is irrelevant.

The fluxes were integrated from 20/3 angles (cf. Sects.~\ref{ss_cif} and
\ref{ss_1d}) for the 3-D/1-D calculations, respectively. For the study
of the CLV, intensities (as a function of $\mu$) were calculated for 11
positions on the Sun ($\mu \equiv {\rm cos}\theta=1.0, 0.9, ..., 0.1, 0.05$)
averaging over 4 directions in $\phi$ and all horizontal (X-Y) 
positions. All 99 snapshots were utilized for the 3-D calculations.

\begin{figure*} [hbtp]
\epsscale{0.98}
\plotone{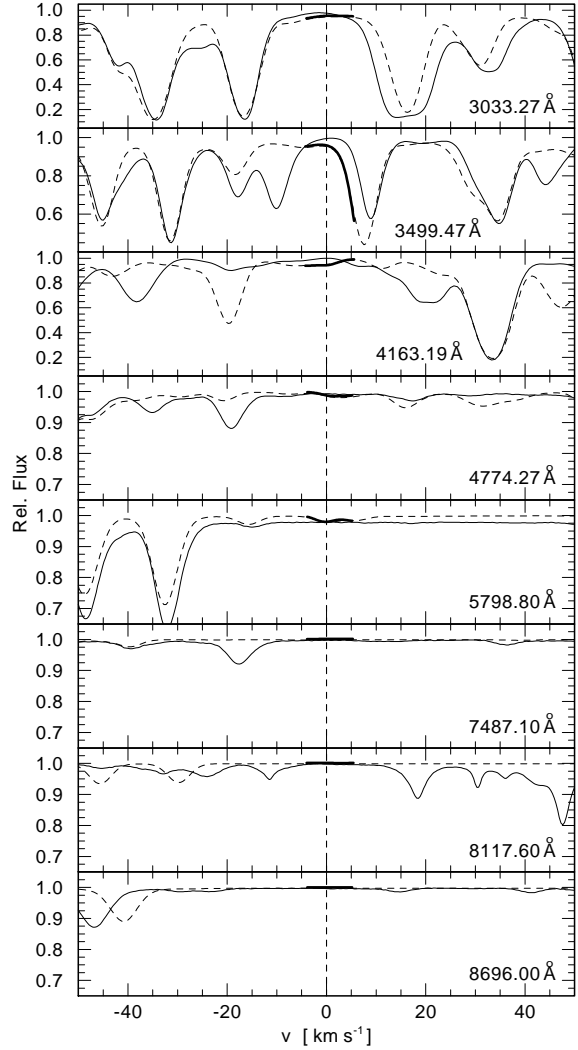}
\caption{
\label{fig_fluxc}
Comparison of the normalized solar spectrum (thin, solid) and the
synthetic spectra from the 1-D Kurucz model (dashed) and the 3-D
Hydro-simulation (thick, solid) for the 8 wavelengths under
consideration.  The 3-D calculations were  performed only for the small 
windows of $\pm 5$ km s$^{-1}$ used here to study the  
center-to-limb variation in the continuum. 
For the normalization, the synthetic spectra were divided
by the corresponding ``pure-continuum'' spectra. The 1-D spectrum has been
convolved with a Gaussian of FWHM$=4.3$ km s$^{-1}$, 
to account for macro-turbulence (FWHM$=4.2$ km s$^{-1}$) and 
the finite resolution of the solar atlas (FWHM$=0.8$ km s$^{-1}$).
Note that the log-$gf$ values
of the individual lines have not been adjusted to match the spectrum.
}
\end{figure*}

The eight frequencies cover a broad spectral range. Although some
neighboring features are poorly matched by our synthetic spectra, the
solar flux spectrum of \cite{kur84} is reproduced well at the
frequencies selected by Neckel \& Labs, and therefore modifications of
our linelist were deemed unnecessary (see Fig.~\ref{fig_fluxc}). The
normalization of the synthetic spectra was achieved by means of
``pure-continuum'' fluxes that were derived from calculations lacking
all atomic and molecular line opacities -- Fig.~\ref{fig_fluxc} shows
that Neckel \& Labs did a superb job selecting continuum windows.

\begin{figure*} [hbtp]
\epsscale{0.94}
\plotone{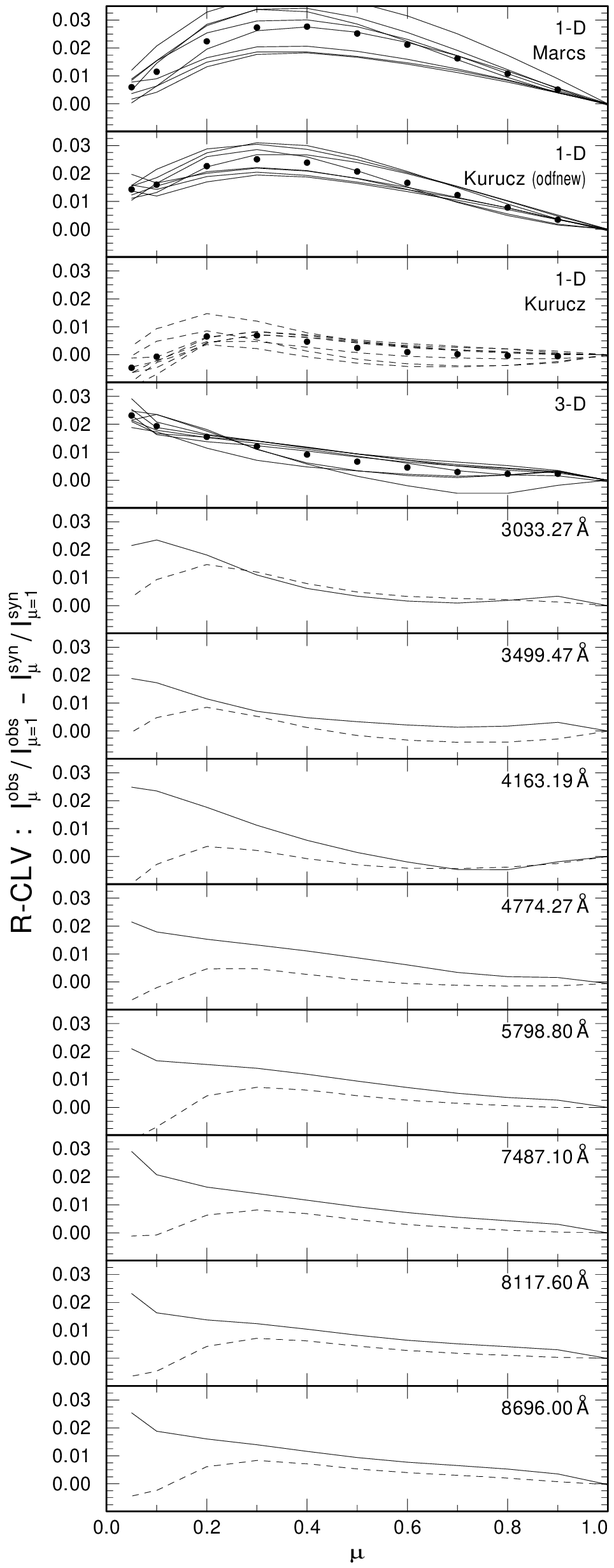}
\caption{
\label{fig_clvc}
Residual CLV's (R-CLV's) in the continuum, i.e. the difference of the
observed and the synthetic normalized CLV's for the 3-D (solid) and the
1-D (dashed) model. The upper four panels show the 1-D and 3-D data
combined, respectively. Average values are indicated by circles. The
lower panels compare separately the data for the 6 wavelengths under
consideration but do not repeat the data from the alternative (ODFNEW) 
Kurucz and the Marcs model. 
Positive R-CLV's indicate that the temperature drops off
too fast in the model atmospheres.
}
\end{figure*}

Comparisons of observed and synthetic CLV's are conducted with datasets
that are normalized with respect to the intensity at the disk center,
i.e. all intensities are divided by the central intensity. We show the
residual CLV's
\begin{equation} 
{\rm R\!-\!CLV} \equiv    I_\mu^{\rm obs}/I_{\mu=1}^{\rm obs} 
        -  I_\mu^{\rm syn}/I_{\mu=1}^{\rm syn},
\end{equation}
\noindent in Fig.~\ref{fig_clvc}. The R-CLV's within each group
are quite homogeneous. In addition to the data derived from our 1-D
Kurucz model (cf. Sect.~\ref{ss_km}), we show also data from two other
1-D models, i.e. a MARCS model (1$^{\rm st}$ panel) from Asplund
(priv. comm.)  and an alternative (odfnew) Kurucz model (2$^{\rm nd}$
panel) from a different model grid
(http://kurucz.harvard.edu/grids.html). The Center-to-Limb variation from
both alternatives show much larger residuals and are not used for the
comparison with the 3-D data. However, the scatter within the 1-D data
demonstrates vividly the divergence that still persist among different
1-D models.

Our reference 1-D model (3$^{\rm rd}$ panel) describes the observed CLV's 
well down to $\mu \approx 0.5$. Closer to the rim the R-CLV's rise
to $\approx 0.1$ at $\mu=0.2$ followed by a sharp decline at the
rim. In 3-D (4$^{\rm th}$ panel) we find on average a linear trend of
the R-CLV's with $\mu$, showing a maximum residual of {\footnotesize
$\gtrsim$} 0.2 close to the rim.

The investigation of the Center-to-Limb variation of the continuum is an
effective tool to probe the continuum forming region at and 
above $\tau \approx 1$. Deviations from the observed CLV's indicate that the
temperature gradient around $\tau \approx 2/3$ is incorrectly reproduced
by the model atmosphere. This can, of course, mean that the gradient in
the model is inaccurate, but it can also signal that the opacity used for
the construction of the model atmosphere differs significantly from the
opacity used for the spectrum synthesis. In that case, the temperature
gradient is tested at the wrong depth due to the shift of the
$\tau$-scales.

Our spectrum calculations suffer from an inconsistency 
introduced by the fact that the abundance pattern and the opacity
cross-sections might differ from what was used when the model was
constructed. In our reference 1-D model, 
we compensate for the new solar iron abundance
($\epsilon_{\rm Fe}: 7.63 \rightarrow 7.45$) and interpolate to 
${\rm log}\,(Z/Z_\sun)=-0.2$ 
in the Kurucz model grid (cf. Sect.~\ref{ss_km}). The 3-D model has been
constructed based on the \citet{gre98} solar abundances
(cf. \citet{asp00}) with $\epsilon_{\rm Fe} = 7.50$ and, to first order,
 no compensation is necessary. (And the same is true for the other two
1-D models considered in Fig. \ref{fig_clvc}.) 
The changes in carbon and oxygen abundances do not
affect the continuum opacities, which are dominated in the optical
by H and H$^-$. Consequently, only metals that
contribute to the electron density 
and therefore to the H$^-$ population (i.e. Fe, Si and Mg) 
are relevant. 


\begin{figure} [hbtp]
\epsscale{0.98}
\plotone{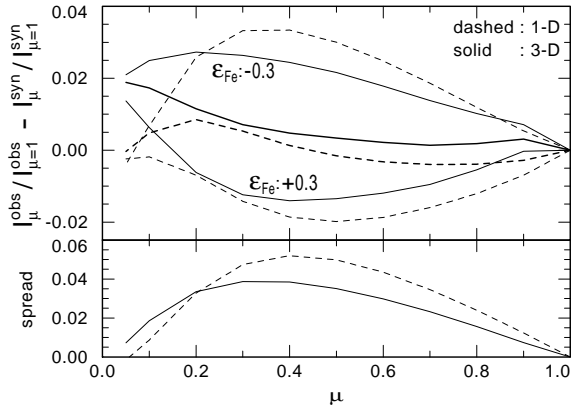}
\caption{
\label{fig_clvcfe}
Upper panel : residual continuum CLV's at 3499.47\,\AA\ derived from the
1-D (dashed) and the 3-D model (solid) with varied Fe abundances of $\pm
0.3$ dex. The unaltered data are highlighted.  Lower panel : Difference
of the data in the upper panel from the calculation with $\epsilon_{\rm
  Fe}=\pm 0.3$ dex.
}
\end{figure}

In order to investigate the
impact of changes of the opacity on the Center-to-Limb
variation 
we have calculated the R-CLV's for the 3-D and our reference 1-D
 models at 3499.47\,\AA\ with two
different Fe abundances ($\pm 0.3$ dex). 
The purpose of the test is to demonstrate the general effect of
opacity variations that can come from different sources, i.e. uncertainties
of abundances and uncertainties of bound-free cross-sections of all relevant
species (not only iron). However, to simplify the procedure we have
modified only the abundance of iron which stands for the
cumulative effect of all uncertainties. In the example the total opacity
is increased by 50\% and decreased by 22\%, respectively. 

Increased opacity, i.e. increased iron abundances, results in
large negative residuals while decreased opacity produces large positive
residuals (Fig.~\ref{fig_clvcfe}). Both models are affected in a similar
way, but the strength of the effect  is
slightly smaller for the 3-D calculation
by about 20\% (cf. Fig.~\ref{fig_clvcfe}, lower
panel). A change in opacity has a significant effect on the CLV
but it does not eliminate the discrepancies.

\begin{figure} [hbtp]
\epsscale{0.98}
\plotone{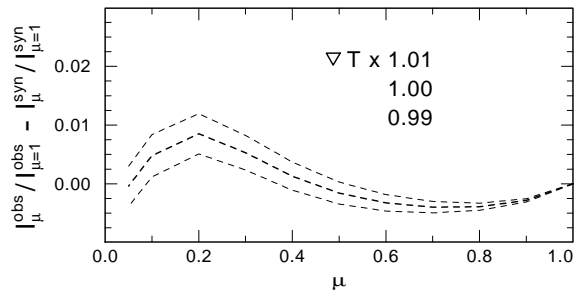}
\caption{
\label{fig_clvct}
Residual continuum CLV's at 3499.47\,\AA\ from the 1-D model with
modified temperature fields. The temperature gradient at $\tau \approx
2/3$ is changed by +1\% (upper curve) and -1\% (lower curve). The
unaltered data are highlighted. 
}
\end{figure}

To estimate the effect of a varied temperature 
gradient on the R-CLV's we have calculated the CLV at
3499.47\,\AA\ for two artificially modified 1-D models
(Fig.~\ref{fig_clvct}). The temperature structure around $\tau_{\rm
Ross}=2/3$ is changed such that the gradient in temperature is increased
and decreased by 1\%, respectively. At $\mu=0.2$, i.e. the position of
the largest discrepancy, the residual of $0.0085$ is changed by $\approx
0.0035$, i.e. by roughly 1/3, indicating a maximum error of the 1-D
temperature gradient of about 3\%. Again, a simple change does not 
lead to perfect agreement, especially when more than one frequency is
considered. 

\begin{figure*} [ht!]
\epsscale{1.6}
\plotone{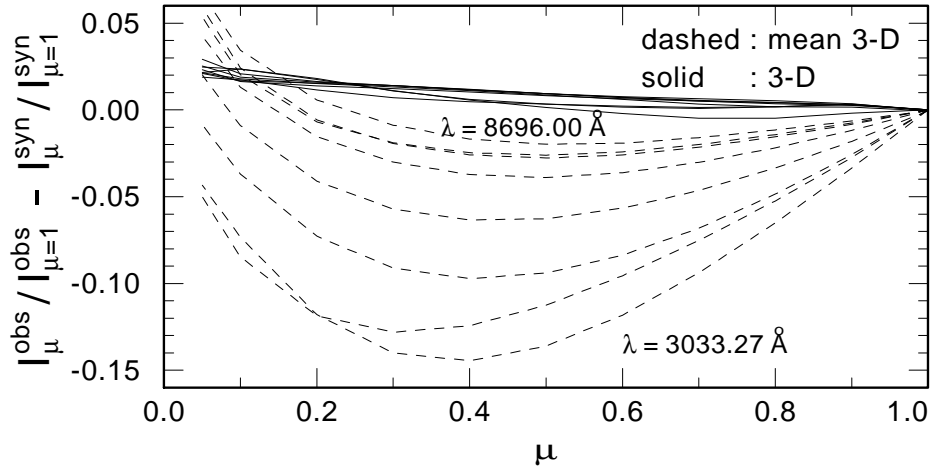}
\caption{
\label{fig_clvm3d}
Residual CLV's in the continuum, i.e. the difference of the observed and
the synthetic normalized CLV's for the 3-D model (solid) and its 1-D
approximation (dashed). In comparison, the approximation in 1-D,
i.e. the 'horizontal' time average of the 3-D model, fails to reproduce
the observed Center-to-Limb variation at all. The maximum absolute
residual is up to 5 times larger, i.e. -0.15 (1-D) vs. 0.03 (3-D).
}
\end{figure*}

Finally, we compare the CLV in the continuum for the average
('horizontal' and over time) 3-D model with the exact data derived
from the radiation transfer in 3-D. Fig.~\ref{fig_clvm3d} shows the
residual CLV's for both models. The discrepancies with the observations
are much more severe for the average 3-D model and it becomes obvious
that it does not represent the original 3-D time series at all. 
Although a 1-D representation would obviously be
highly desirable because it would allow to quickly calculate spectra by
means of a 1-D radiation transfer code, 
this turns out to be a very poor approximation in this case.

\citet{ayr06} have carefully investigated the rotational-vibrational bands
of carbon monoxide (CO) in the solar spectrum and have derived oxygen
abundances from three models, i.e. the Fal C model \citep{fon93}, a 1-D
model that is especially adapted to match the Center-to-Limb variation
of the CO-bands (COmosphere), and from the averaged 3-D time series. In
all three cases, temperature fluctuations are accounted for in a
so-called 1.5-D approximation, in which profiles from 5 different
temperature structures are averaged. By assuming a C/O ratio of 0.5, 
\citet{ayr06} derive a high oxygen abundance close to the ``old'' value
from \citet{gre98} from both the Fal C and the COmosphere
model, discarding the low oxygen abundance derived from the
mean 3-D model because its temperature gradient is too steep around
$\tau_{\rm 0.5\,\mu m}\approx 1$ and fails to reproduce the observed
Center-to-Limb variations. 

Our current study documents that the mean 3-D
model is not a valid approximation of the 3-D time series, and therefore
its performance cannot be taken as indicative of the performance of the
3-D model, and in particular of its temperature profile.
We find that the Center-to-Limb variation of the continuum 
predicted by the 3-D simulation 
matches reasonably well (i.e. similar to the best 1-D model in our study) 
the observations. The results by Scott et al., based on 3-D
radiative transfer on the same hydrodynamical simulations used here, 
indicate that the
observed CO ro-vibrational lines are consistent with the low oxygen
and carbon abundances. Our results show that there is no reason to distrust
the 3-D-based abundances on the basis of the simulations having a wrong thermal 
profile.

\subsection{Lines}
\label{s_lines}

We study the Center-to-Limb variation of a number of lines by comparing
observations of the quiet Sun taken at 6 different heliocentric angles
to synthetic profiles derived from 3-D and 1-D models. The observations
are described in detail by \citet{allII} and were previously 
used for the investigation of inelastic 
collisions with neutral hydrogen on oxygen lines formed
under non-LTE conditions\footnote{Data available at
http://hebe.as.utexas.edu/izana}. The observations cover 8 spectral
regions obtained at 6 different positions on the Sun. The first 5 slit 
positions are centered at heliocentric angles of 
$\mu \equiv {\rm cos}\,\theta=$
1.00, 0.97, 0.87, 0.71 and 0.50. The last position varies between $\mu=$
0.26 and 0.17 for different wavelength regions. This
translates to distances of the slit center from the limb of the Sun in
arc\,min of 16.00', 12.11', 8.11', 4.74', 2.14', 0.54' and 0.24', 
assuming a diameter of the Sun of 31.99'.  For both of these last
positions the slit extends beyond the solar disk and the center of the
illuminated slit corresponds to $\mu=$ 0.34 and 0.31 (0.96' and 0.78').

We have calculated a variety of line profiles for the 6 positions
defined by the center (in $\mu$) of the illuminated slit. Although the
slit length, 160\,arcsec, is rather large, test calculations show that
averaging the spectrum from six discrete $\mu$-angles 
spanning the slit length gives virtually the
same equivalent width than the spectrum from the central $\mu$. For
$\mu=0.5$, the second last angle, the difference amounts to a marginal
change of the log-$gf$ value of about 0.01.
To further reduce the computational burden  we have derived the
average 3-D profiles from calculations taking only 50 (every other) of the 99
snapshots into account.

We have selected 10 seemingly unblended lines from 5 different
neutral ions. The list of lines is compiled in Table~\ref{tabI}. The
log-$gf$ values for most lines were adopted from laboratory measurements
at Oxford (e.g. \citet{bla95} and references therein) and by
\citet{obr91}. 


\begin{table*} [hbtp]
\begin{center}
\caption{Lines \label{tabI}}
\vspace{1mm}
\begin{tabular}{cccccccc}
\tableline
\tableline
Ion& $\lambda$& $R'$& max($\theta$)& log-$gf$ 
& log $\Gamma_{\rm Rad}$ & log $\Gamma_{\rm Stark}$ & log $\Gamma_{\rm VdW}$\\
   &     [\AA]&    \tablenotemark{(a)} &         [deg]&   &
    \tablenotemark{(b)}  & \tablenotemark{(c)} &   \tablenotemark{(d)}   \\
\tableline
Fe\,I& 5242.5&  56000& 75& -0.970& 7.76&  -6.33&  -7.58\\
Fe\,I& 5243.8&  56000& 75& -1.050& 8.32&  -4.61&  -7.22\\
Fe\,I& 5247.0&  56000& 75& -4.946& 3.89&  -6.33&  -7.82\\
Fe\,I\tablenotemark{(e)}& 6170.5&  77000& 80& -0.380& 8.24&  -5.59&  -7.12\\
Fe\,I\tablenotemark{(f)}& 6200.3& 206000& 75& -2.437& 8.01&  -6.11&  -7.59\\
Fe\,I& 7583.8& 176000& 80& -1.880& 8.01&  -6.33&  -7.57\\
\tableline
Cr\,I& 5247.6&  56000& 75& -1.627& 7.72&  -6.12&  -7.62\\
Ni\,I& 5248.4&  56000& 75& -2.426& 7.92&  -4.64&  -7.76\\
Si\,I& 6125.0&  77000& 80& -0.930& \nodata \tablenotemark{(g)}&   \nodata&   \nodata \\
Ti\,I& 6126.2&  77000& 80& -1.425& 6.85&  -6.35&  -7.73\\
\tableline
\end{tabular}
\tablenotetext{a}{cf. \citet{allII}, $R'$ is the resolving power 
measured relative to the FTS ($R_{\rm FTS} \approx 400\,000$) spectrum
at the center disk provided by \citet{braI}. }
\tablenotetext{b}{$\Gamma = \gamma$, where $\gamma$ 
is the damping constant (FWHM of a Lorentzian profile, see,
e.g. Eq. (11.13) in Gray 1992),
in rad s$^{-1}$.}
\tablenotetext{c}{$\Gamma= \gamma/{\rm Ne}$, where Ne indicates the 
 number density of electrons at a temperature of 10,000 K (cgs units).}
\tablenotetext{d}{$\Gamma= \gamma/{\rm N_H}$, where N$_{\rm H}$  
is the hydrogen number density at a temperature of 10,000 K.}
\tablenotetext{e}{noticeably blend at +5\,km\,s$^{-1}$}
\tablenotetext{f}{marginal blend at +5\,km\,s$^{-1}$}
\tablenotetext{g}{Approximate values were adopted for this line, see, e.g. Gray 1992}
\end{center}
\end{table*}

\begin{figure} [hbtp]
\epsscale{0.98}
\plotone{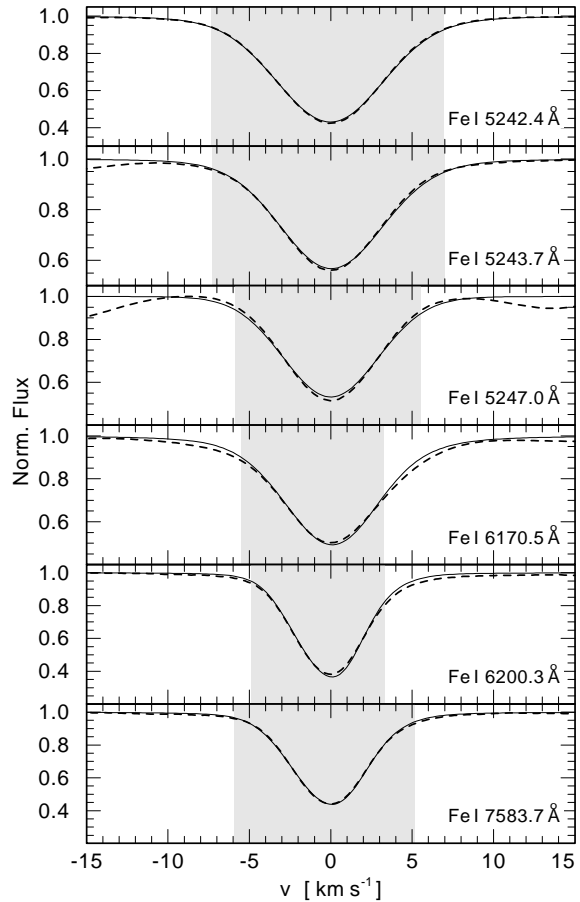}
\caption{
\label{fig_l1}
Iron lines under consideration. Observed (dashed) and synthetic (solid)
profiles are shown for the  disk center. The grey areas mark the
velocity ranges used for the calculation of the line equivalent
widths. The zero of the velocity scale refers always to the center of
the observed line profile, as approximately 
determined by polynomial fitting. The log-$gf$ values are modified to match the observed line
equivalent widths. For most lines the profiles match well. However, the
synthetic profile of the line at 5247.0\,\AA\ seems to be broader than
the observed profile. The line at 6200.3\,\AA\ is marginally blended
around +5\,km\,s$^{-1}$. The line at 6170.5\,\AA\ is noticeable blended around
+5\,km\,s$^{-1}$. For both lines the wavelength interval is decreased
accordingly.
}
\end{figure}

\begin{figure} [hbtp]
\epsscale{0.98}
\plotone{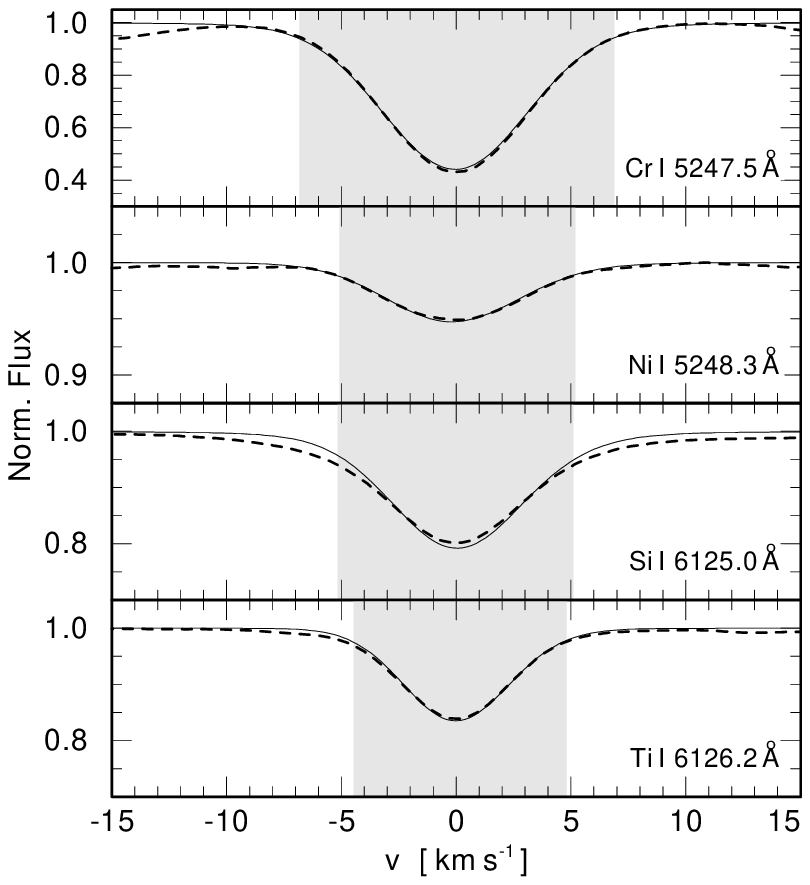}
\caption{
\label{fig_l2}
Non-iron lines under consideration. Observed (dashed) and synthetic
(solid) profiles are shown 
for the center disk. The grey areas mark the wavelength
ranges used for the calculation of the line equivalent widths. The
log-$gf$ values are modified to match the observed line equivalent
widths. For most lines the profiles match well. However, the
synthetic profile of the Si\,I line at 6125.0\,\AA\ is noticeably
narrower than observed.
}
\end{figure}

We are interested in how synthetic lines profiles deviate from
observations as a function of the position angle $\mu$ for two
reasons. First of all, any clear trend with $\mu$ would reveal
shortcomings of the theoretical model atmospheres similar to our
findings presented in Sect.~\ref{ss_continuum}. But, arguably, even more
relevant is the fact that any significant deviation (scatter) would add
to the error bar attached to a line-based abundance determination.

In our present study we compare synthetic line profiles from 3-D and 1-D
models with the observations. Due to the inherent deficiencies of the
latter models, i.e. no velocity fields and correspondingly narrow and
symmetric line profiles, etc., we focus on line strengths and compare
observed and synthetic line equivalent widths, rather than comparing the
line profiles in detail. To be able to detect weak deviations, we have devised
the following strategy. We have identified wavelength intervals around
each line under consideration for the contribution to the line
equivalent widths and have calculated series of synthetic line profiles
in 1-D and 3-D with varied log-$gf$ values that encircle the
observations with respect to their equivalent widths. That allowed us to
determine by interpolation the log-$gf$ value required to match the
observed line equivalent widths separately for each position angle
(``Best-Fit''). To keep interpolation errors at a marginal level we have
applied a small step of $\Delta$(log-$gf)=0.05$ for these series of
calculations. A simple normalization scheme has been applied. All
profiles have been divided by the maximum intensity found in the vicinity
  of the line center(within $\pm15\,{\rm km\,s^{-1}}$). 
We convolved the synthetic profiles with a Gaussian as to mimic the
instrumental profile (see Table 1).
An additional Gaussian broadening is applied
to the line profiles from the 1-D calculation to account for 
macro-turbulence; this value was
adjusted for each line in order to reproduce the  line profiles observed
at the disk center.

Finally we have translated variations of line strength into variations
of abundance, i.e. we have identified $\Delta$ log-($gf$) =
$\Delta$log-$\epsilon$. This approximation is valid because the impact
of slight changes in a metal abundance on the continuum in the optical
is marginal. Note that it is not the intent of this study to derive
metal abundances from individual lines. Such an endeavor would require a
more careful consideration regarding line blends, continuum
normalization, and non-LTE effects. 

All calculations described in this section are single-line calculations,
i.e. no blends with atomic or molecular lines are accounted for.
The observations did not
have information on the absolute wavelength scale (see Allende 
Prieto et al. 2004), but that is not important for our purposes and
the velocity scales in Figs. \ref{fig_l1} and \ref{fig_l2} are relative
to center of the line profiles. The
individual synthetic profiles were
convolved with a Gaussian profile to match
the observed profiles (cf. Table~\ref{tabI}).
We were generally able to achieve a better fit of the
observations when slightly less broadening was applied to the 3-D
profiles (0.3\% in case of Fe\,I 5242.5). Since we know from previous
investigations that the theoretical profiles
derived from 3-D Hydro-models match the observations well, we argue
that the resolution of the observations is actually slightly higher than
estimated by \citet{allII}. An alternative explanation would be that the
amplitude of the velocity field in the models is too high. Such a
finding, if confirmed, deserves a deeper investigation but is beyond the
scope of this study since line equivalent widths are only marginally (if
at all) affected.

\begin{figure} [hbtp]
\epsscale{0.98}
\plotone{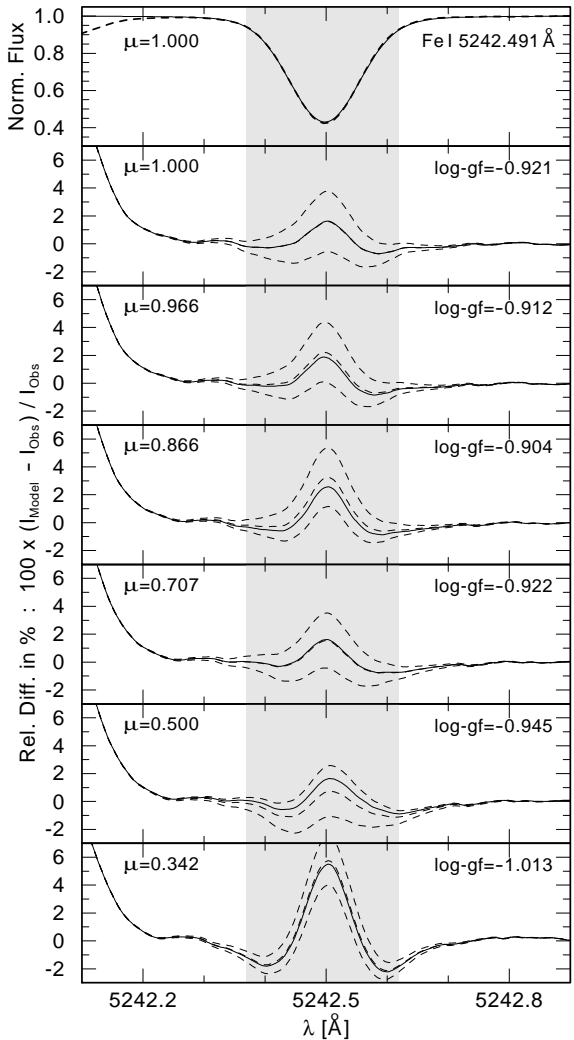}
\caption{
\label{fig_lex}
Fit of the Fe\,I line at 5242.5\,\AA\ with synthetic profiles derived
from the 3-D Hydro model. Upper panel: Normalized profiles of the center
disk observation (dashed) and the ``Best-Fit'' synthetic profile
(solid). Lower panels: Relative difference of the observation and three
synthetic profiles (dashed), i.e. three different ${\rm log}\,gf$-values 
for the 6 angles under consideration. The equivalent widths of the
``Best-Fits'' (solid) match the observed equivalent widths. The
wavelength range considered for the equivalent width is highlighted in
grey. The fit improves when a slightly higher resolution (by $\approx
0.3\%$) is assumed. However, line equivalent widths are only marginally
affected.
}
\end{figure}

We introduce the lines under consideration by showing the observed
center-disk line profiles and the ``Best-Fits'' derived from the 3-D
calculations of the six Fe\,I lines and the four lines from other ions in
Figs.~\ref{fig_l1} and \ref{fig_l2}, respectively. 
In Fig.~\ref{fig_lex} we exemplify the fitting process by means of the Fe\,I
line at 5242.5\,\AA\ and show the relative difference between the
observation and a variety of model calculations for all 6 angles under
consideration. The ``Best-Fit'' log-$gf$ values are derived by
interpolation to match the observed equivalent widths from the spectral
region around the line profile.


\begin{figure*} [hbtp]
\epsscale{1.3}
\plotone{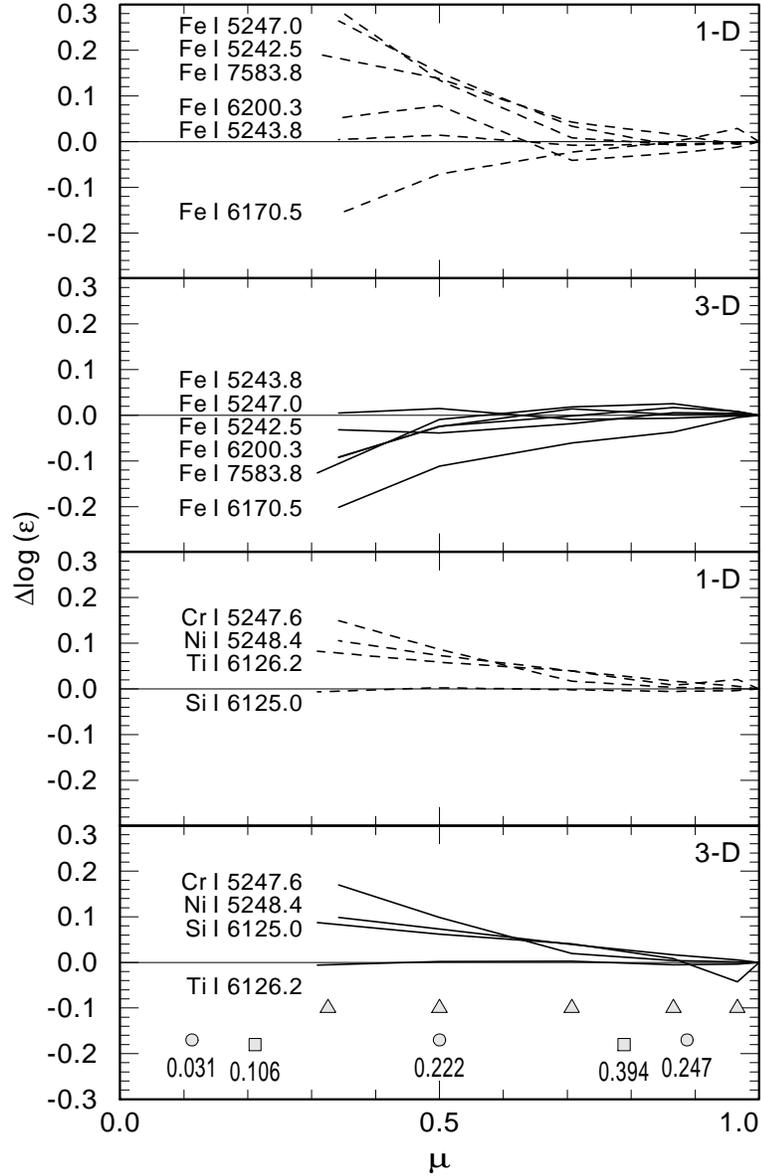}
\caption{
\label{fig_clvl}
Relative abundance variations with respect to the center disk
($\mu\!\!=\!\!1$) from the 1-D (dashed) and the 3-D calculations,
respectively. The upper two panels show the six Fe\,I lines, the lower two
panels show the four lines from the other elements. Line designations are
given in the left part of the plot. In the second panel from the top,
the curves for the iron lines at 5242.5 and 6200.3 \AA\ 
overlap with each other for the two positions closest to the limb.
The bottom panel also indicates the
six $\mu$-angles of the observations (triangles, $\mu\!\!=\!\!1$ not
shown, shallowest angles varies slightly) and the $\mu$-angles used in a
2-point (squares) and a 3-point (circles) Gaussian flux integration and
their respective integration weights ({\footnotesize $\sum$} $w_\mu =
0.5$). 
}
\end{figure*}

We have obtained ``Best-Fits'' for all 10 lines (cf. Table~\ref{tabI})
and present the log-$gf$ values as a function of $\mu$ in
Fig.~\ref{fig_clvl}. Be reminded that the aim of this
 study is not the measurement of absolute abundances: 
we focus on relative numbers and normalize our results with
respect to the  disk center ($\mu=1$).

For improved readability we subdivide our findings presented in
Fig.~\ref{fig_clvl} into 4 distinct groups, i.e. iron/non-iron lines and
1-D/3-D calculations, respectively.  We focus our discussion on
the first five data points because we have some indications that the data
obtained for the shallowest angle is less trustworthy than the data
from the other angles: {\it i}) the relative contribution 
of scattered light was estimated 
from the comparison of the center-disk spectrum with the FTS
spectrum taken from \citet{braI}, and the outer-most position was the
only one for which the entire slit was not illuminated, 
{\it ii}) for all 10 lines the fit of the line profiles
for this particular angle is the worst (cf. Fig.~\ref{fig_lex}) and {\it
iii}) the scatter in our data presented in Fig.~\ref{fig_clvl} is the
largest for this angle. 
Fortunately, the flux integration is naturally 
biased towards the center of the disk.

We find this systematic behavior for all 6 iron lines: 
$\Delta{\rm log} (\epsilon)$ is larger or equal in 1-D compared to 3-D,
for all but one line (Fe\,I 6170.5\,\AA) $\Delta{\rm log}
(\epsilon)$ is positive or zero for the 1-D calculations, and 
$\Delta{\rm log} (\epsilon)$ is negative or zero for all 3-D
calculations.
The Fe\,I line at 6170.5\,\AA\ stands out in both comparisons. In 1-D it
is the only line with a negative $\Delta{\rm log} (\epsilon)$ and in 3-D
it shows the by far largest negative $\Delta{\rm log} (\epsilon)$. This
might be related to the noticeable line blend 
(cf. Fig.~\ref{fig_l1}).

The iron lines calculated in 3-D indicate a uniform trend of decreased
log-$gf$ values with increased distance from the center-disk.  The
average decrease at $\mu=0.5$ for this group is -0.015 (Fe\,I
6170.5\,\AA\ excluded). From the 1-D calculations we derive the
opposite trend for the same group of lines and obtain an average of
0.103. Obviously, the 3-D model performs significantly better than the 1-D
reference model regarding the center-to-limb variation of Fe I lines,
even when equivalent widths,  and not line asymmetries or shifts, are
considered.

For these five Fe\,I lines we obtain an average difference (1-D
vs. 3-D) of 0.12 at $\mu=0.5$. To estimate the impact on abundance
determinations based on solar fluxes we apply 
a 3-point Gaussian integration, 
neglecting the shallowest angle at $\mu=0.11$ (which has, by far, the
smallest integration weight) for which we have no data, and 
assuming that the good agreement between the 1-D
and the 3-D calculations for the central ray implies an equally good 
agreement for the first angle at $\mu=0.89$. 
These estimates lead to an abundance correction of
approximately 0.06 dex 
between 1-D and 3-D models due to their different center-to-limb variation.
\citet{asp00} found a similar correction from the comparison
of 1-D and 3-D line profiles at the disk center.

For the 4 non-iron lines we find a uniform trend of increasing log-$gf$
values with decreasing $\mu$ for both, the 1-D and the 3-D dataset.  The
systematic behavior is similar to what we find for the iron lines, but 
now the performance of the 1-D and 3-D models is similar, and the offsets
are in the same sense: larger abundances would be found towards the limb
for both models.

\section{Conclusion}

The photosphere of cool stars and the Sun can be  described
by stellar atmospheres in 1-D and 3-D. Since the 3-D models add more
realistic physics, i.e. the hydrodynamic description of the gas, they
can be seen truly as an advancement over the 1-D
models. However, this refinement increases the computational effort by
many orders of magnitude. In fact, the computational workload becomes so
demanding, that the description of the radiation field has to be cut
back to very few frequencies, i.e. to a rudimentary level that had been
surpassed by 1-D models over 30 years ago. 
Overall we are left with the
astonishing situation that a stellar photosphere can be modeled by
either an accurate description of the radiation field with the help of a
makeshift account of stellar convection (Mixing-length theory), or by an
accurate description of the hydrodynamic properties augmented by a
rudimentary account of the radiation field.

It is evident that individual line profiles can be described to a
much higher degree and without any artificial micro- or 
macro-turbulence by the 3-D
Hydro models, as the simulations account for Doppler-shifts from
differential motions within the atmosphere. We know from detailed 
investigations of line profiles that
the velocity field is described quite accurately and that
the residuals of the fittings to line
profiles are reduced by about a factor of 10. However, it is not 
obvious how the 3-D models compare to their 1-D counterparts
when it comes to reproduce spectral energy distributions and line strengths.

We study the solar Center-to-Limb variation for several lines and continua, to
probe the temperature structure of 3-D models. The work is facilitated
by the new code {\sc Ass$\epsilon$t}, which allows for the fast and
accurate calculation of spectra from 3-D structures. In
comparisons to other programs (e.g. \citet{asp00, lud07}), the attributes
of the new code are a greater versatility, i.e. the ability to handle
arbitrarily complicated lines blends on top of non-constant background
opacities, higher accuracy due to the proper incorporation of scattering
and improved (higher-order) interpolation schemes, and a higher 
computational speed.

In our study we find that regarding center-to-limb variations, 
the overall shortcomings of the 3-D model are
roughly comparable to the shortcomings of the 1-D models. Firstly, we
conclude from the investigation of the continuum layers that the models'
temperature gradient is too steep around $\tau\approx2/3$. This behavior
is more pronounced for the 3-D model which shows a drop in intensity
(with $\mu$) that is about twice the size of the drop displayed by our
reference 1-D model, but at the same time smaller than the discrepancies 
found for two other (newer!) 1-D structures.
Secondly, the line profiles for different position angles on the
Sun cannot be reproduced by a single abundance. For Fe I lines, the
abundance variation between the disk center and $\mu=0.5$ is about 0.1 dex
for our reference 1-D model, but only $0.015$ dex (and with the opposite sign)  
for the 3-D simulations, albeit the calculations for lines of other neutral
species suggest a more balanced outcome.

Overall we conclude that the 1-D and the 3-D models match the observed
temperature structure to a similar degree of accuracy. This is somewhat
surprising but it might be that the improved description of the
convective energy transport is offset by deficiencies introduced by the
poor radiation transfer. Once new Hydro models based on an upgraded
radiation transfer scheme (i.e. more frequencies and angles, better
frequency binning) become available in the near future (Asplund,
priv. comm.), we will be able to test this hypothesis. 
It will become clear whether  
focusing on refining the radiation transfer will be enough to achieve better
agreement with  observations, or the hydrodynamics  needs
to be improved as well.

\acknowledgments

We thank M. Asplund for providing us with the 3-D hydrodynamical simulation and
the 1-D MARCS model, and M. Bautista, I. Hubeny, and S. Nahar 
for crucial assistance computing opacities.
We extent our thanks to the late John Bahcall, Andy Davis, and Marc 
Pinsonneault for their interest on our tests of the solar simulations, 
which enhanced our motivation to carry out this work.
Continuing support from NSF (AST-0086321),  NASA (NAG5-13057 and NAG5-13147), 
and the Welch Foundation of Houston is greatly appreciated.



\end{document}